\documentclass[conference]{IEEEtran}

\usepackage{amsfonts}
\usepackage{amssymb}
\usepackage{stfloats}
\usepackage{cite}
\usepackage[pdftex]{graphicx}
  \graphicspath{{./figures/}}
  \DeclareGraphicsExtensions{.pdf}
\usepackage{amsmath}
\usepackage{dsfont}
\usepackage{array}
\usepackage{url}
\usepackage{booktabs}
\usepackage{multirow}
\usepackage{psfrag}
\usepackage{subfigure}
\usepackage{algorithm}
\usepackage{algpseudocode}
\usepackage[affil-it]{authblk}
\usepackage[english]{babel}
\usepackage{blindtext}
\usepackage{url}
\usepackage{epstopdf}
\usepackage{verbatim}
\usepackage{xcolor}

\newtheorem{Prob}{Problem}

\title{Efficient MIMO Detection with Imperfect Channel Knowledge - A Deep Learning Approach}

\author{Qian Chen$^{\dagger}$, Shunqing Zhang$^{\dagger}$, Shugong Xu$^{\dagger}$ and Shan Cao$^{\dagger}$\\
$^{\dagger}$ Shanghai Institute for Advanced Communication and Data Science, \\
Key laboratory of Specialty Fiber Optics and Optical Access Networks, \\
Shanghai University, Shanghai, 200444, China\\
Email:\{chenqian, shunqing, shugong, cshan\}@shu.edu.cn}

\begin{document}

\maketitle

\begin{abstract}
Multiple-input multiple-output (MIMO) system is the key technology for long term evolution (LTE) and 5G. The information detection problem at the receiver side is in general difficult due to the imbalance of decoding complexity and decoding accuracy within conventional methods. Hence, a deep learning based efficient MIMO detection approach is proposed in this paper. In our work, we use a neural network to directly get a mapping function of received signals, channel matrix and transmitted bit streams. Then, we compare the end-to-end approach using deep learning with the conventional methods in possession of perfect channel knowledge and imperfect channel knowledge. Simulation results show that our method presents a better trade-off in the performance for accuracy versus decoding complexity. At the same time, better robustness can be achieved in condition of imperfect channel knowledge compared with conventional algorithms. 
\end{abstract}


\section{Introduction}%
MIMO systems \cite{telatar1999capacity}, proposed by Telatar 20 years ago, have been selected as one of the key features in the current LTE systems and the massive deployment of MIMO systems \cite{massiveMIMO} are often regarded as a breakthrough for 5G systems. Although the transmit side processing for MIMO transmission is often straight forward, the information detection problem at the receiver side is in general difficult \cite{yang2015fifty}. 

Traditional MIMO detection methods include linear and non-linear approaches. For example, linear MIMO detector includes zero forcing (ZF) \cite{ZF} or minimum mean square error (MMSE) \cite{MMSE} equalizers, while non-linear MIMO detector often rely on minimum distance based maximum likelihood (ML) detection \cite{ML}. Although the ML-type decoding provides optimal detection performance in theory, the associated decoding complexity is in general unaffordable with current technology \cite{kim2009new}. In order to provide a better tradeoff, sphere decoding \cite{sphere} and successive interference cancellation \cite{MMSE-SIC}\cite{ZF-SIC} are often applied in practice.

Recently, with the development of machine learning, more complicated problems in wireless communications can be formulated and efficiently solved by this framework. In the physical layer process, for instance, \cite{liang2018iterative} proposes an iterative brief propagation - convolutional neural network architecture for channel decoding, and variational auto-encoder has been applied for blind channel equalization \cite{caciularu2018blind}. Moreover, the authors in \cite{DeepMIMOdetetion} proposed DetNet and show that the gradient projection based detection algorithm can be well approximated by a single layer neural network and the corresponding decoding performance is computationally inexpensive with near semidefinite relaxation (SDR) detection accuracy. Although the above machine learning based approaches show promising gain over the traditional signal detection methods, the following issues in the conventional MIMO detectors have {\em not} been addressed based on our current investigation.

\begin{itemize}
    \item {\em Robust Detection with Imperfect Channel Knowledge} One of the key issues in the MIMO detection design lies in the imperfectness of channel knowledge due to the practical channel estimation method, the time-varying nature of communication devices and limited number of reference signals \cite{ZF}. Since the deep learning framework has powerful generalization ability with respect to the input datasets, a straight forward question is {\em whether we can rely on this generalization ability to handle the robust detection problem with imperfect channel knowledge}. 
    \item {\em Efficient Deep Learning Framework for Detection} Another issue that has not been solved is the efficient deep learning network architecture for the MIMO detection. Although the auto-encoder based solution has been proposed in \cite{yan2017signal} to jointly combine the equalizer and demodulator, the associated processing complexity is still significant if compared with MMSE or ZF based detection. Therefore, an efficient learning framework with only receiver side knowledge will be desirable.
\end{itemize}

In this paper, we exploit the generalization capability of neural networks to address the robust MIMO detection problem as illustrated before. To be more specific, we test convolutional neural network (CNN) and deep neural network (DNN) based approach to model the non-linear transfer function between MIMO transmitters and receivers, the result show DNN-based approach has better BER performance and lower complexity then CNN-based method. Through numerical examples, we show that the proposed DNN-based method outperforms the conventional ZF or MMSE based schemes by using the neural network to directly get a mapping from end to end of the bit stream. Meanwhile, DNN-based solution provides a better tradeoff between the decoding performance and the decoding complexity, which may pave the way for future efficient detection design.

The rest of this paper is organized as follows. In Section~\ref{sect:system}, we briefly introduce the background of MIMO detection problem and formulate the corresponding mathematical problems in Section~\ref{sect:method}. The deep learning method based MIMO detection framework is discussed in Section~\ref{sect:frame} and the numerical results are shown in Section~\ref{sect:experiment}. Finally, concluding remarks are given in Section~\ref{sect:conc}.

\section{System Model} \label{sect:system}
\begin{figure*}[htbp]
\centering
        \includegraphics[width=5.5in]{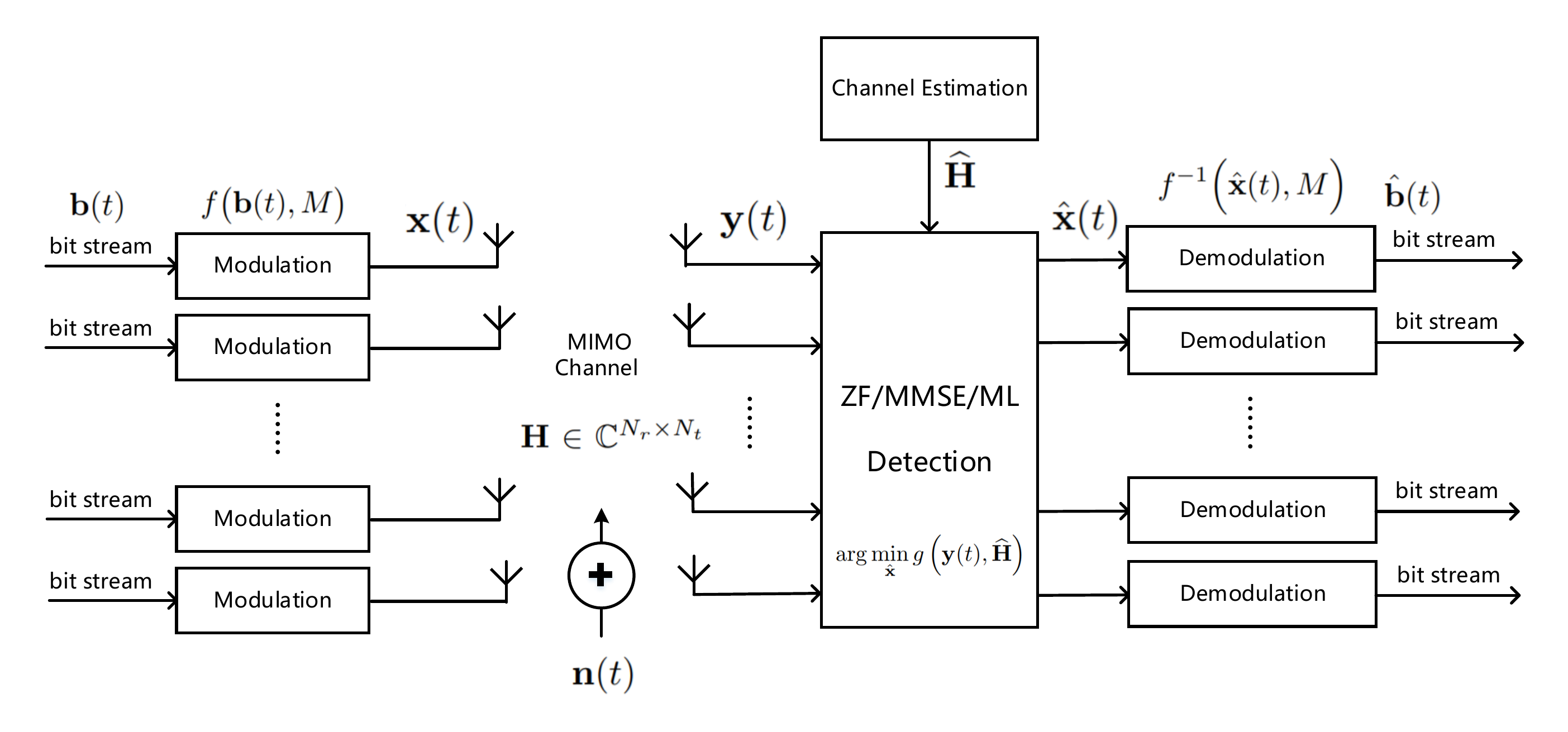}
            \caption{An overview of MIMO system. In this system, the bit streams are sent to the transmitting antenna after modulation. Then, they are transmitted through MIMO channel which contains interference and are finally received by the receiving antenna.}
        \label{fig:overview}
\end{figure*}
Consider a downlink MIMO transmission as shown in Fig.~\ref{fig:overview}, where $N_t$ transmit antennas are delivering messages to $N_r$ antennas using vertical Bell Labs layered space-time (V-BLAST) technique \cite{wei2014power}. Given the time slot $t$ and the modulation size $2^{M}$, a binary information stream $\mathbf{b}(t) \in \mathbb{B}^{(M \times N_t)\times 1}$ are modulated through function $f(\cdot, M)$ and the modulated symbols, $\mathbf{x}(t) \in \mathbb{C}^{N_t \times 1}$, are given by  $\mathbf{x}(t) = f\big(\mathbf{b}(t), M\big)$. Denote $\mathbf{y}(t) \in \mathbb{C}^{N_r \times 1}$ to be the received symbols and the mathematical model for $\mathbf{y}(t)$ are given by,
\begin{eqnarray}
\mathbf{y}(t) = \mathbf{H} \mathbf{x}(t) + \mathbf{n}(t)
\end{eqnarray}
where $\mathbf{H} \in \mathbb{C}^{N_r \times N_t}$ denotes the flat Rayleigh fading coefficients among transmit and receive antenna pairs and $\mathbf{n}(t) \in \mathbb{C}^{N_r \times 1}$ is the additive white Gaussian noise with zero mean and unit variance, i.e. $\mathbf{n}(t) \sim \mathbb{CN}(\mathbf{0},\mathbf{I}_{N_r})$.

In the practical systems, due to the limited power and resources for pilot symbols, the estimated channel state information (CSI) at the receiver side can not be perfect in general. For illustration purpose, we denote $\widehat{\mathbf{H}}$ to be the imperfect CSI at the receiver side and the mathematical expression is given by \cite{li2018accurate}, 
\begin{eqnarray}
\widehat{\mathbf{H}} = \mathbf{H} + \Delta \mathbf{H}
\end{eqnarray}
where $\Delta \mathbf{H}$ denotes the estimation errors. If maximum likelihood (ML) based channel estimation mechanism is applied, $\Delta \mathbf{H}$ can be modeled through independent identical distributed (i.i.d.) complex Gaussian distribution with zero mean and scaled identity covariance matrix \cite{li2018accurate}, i.e., $\Delta \mathbf{H} \sim \mathbb{CN}(\mathbf{0},\sigma_{e}^2\mathbf{I}_{N_r})$, where ${\sigma_e}^2=\frac{N_t}{N_p \cdot E_p}$ with $N_p$ and $E_p$ representing the number and the power of pilot symbols respectively.

With the imperfect CSI $\widehat{\mathbf{H}}$ and the observed symbols $\mathbf{y}(t)$, the detection process at the receiver side can be modeled through, 
\begin{eqnarray}\label{eqn:sym_det}
\hat{\mathbf{x}}(t) = \arg \min_{\mathbf{x}(t)} g\left(\mathbf{y}(t), \widehat{\mathbf{H}}\right) 
\end{eqnarray}
where $\hat{\mathbf{x}}$ is selected from all the possible modulated symbols. By applying the inverse of modulation function, the detected bits for time slot $t$ is given by,
\begin{eqnarray}
\hat{\mathbf{b}}(t) = f^{-1}\Big(\hat{\mathbf{x}}(t), M\Big) \label{eqn:demo}
\end{eqnarray}

The following assumption is made through the rest of this paper. We consider block channel fading assumption, where the channel condition $\mathbf{H}$ remain static within each period $T$ and varies independently among different fading periods. 

\section{Problem Formulation and Traditional Approach}\label{sect:method}

In this section, we formulate the bit error rate (BER) minimization through a general optimization framework. With the above mathematical models of the decoding process, we can write the BER minimization problem as follows.
\begin{Prob}[BER Minimization] \label{prob:BER}
\begin{eqnarray}
&\underset{\hat{\mathbf{b}}(t)}{\textrm{minimize}} & \lim_{N \to \infty }\frac{1}{N}\sum\limits_{n=1}^N \sum\limits_{t=(n-1)T+1}^{nT}\frac{\parallel \hat{\mathbf{b}}(t)-\mathbf{b}(t) \parallel^{2}}{T \times M \times N_t}\\
& \textrm{subject to} &\mathbf{x}(t) = f\big(\mathbf{b}(t), M\big),\\
&&\mathbf{y}(t) = \mathbf{H}_n \mathbf{x}(t) + \mathbf{n}(t),\nonumber \\
&&\forall t\in \left[(n-1)T+1,nT\right], \\
&& \widehat{\mathbf{H}}_n = \mathbf{H}_n + \Delta \mathbf{H}, \quad \widehat{\mathbf{H}}= \left\{\widehat{\mathbf{H}}_n\right\}, \\
&& \hat{\mathbf{b}}(t) = h \left(\mathbf{y}(t), \widehat{\mathbf{H}}\right).
\end{eqnarray}
where $h(\cdot)$ denotes a combined procedure of symbol detection \eqref{eqn:sym_det} and demodulation \eqref{eqn:demo}, and we add subscripts $n$ to the channel conditions $\widehat{\mathbf{H}}$ and $\mathbf{H}$ to represent the $n^{th}$ fading period.
\end{Prob}

To directly solve Problem~\ref{prob:BER} is in general challenging, since the above formulation contains nonlinear functions, such as $f(\cdot)$ and $h(\cdot)$. In addition, the searching space of $\hat{\mathbf{b}}(t)$ is discrete and the brute force approach to obtain the optimal solution requires exponential complexity. To address this challenge, conventional schemes reformulate Problem~\ref{prob:BER} as the following symbol error rate (SER) minimization problem\footnote{By assuming Gray coding, BER can be well approximated by a linear scaling of SER \cite{ML}, and therefore, Problem~\ref{prob:BER} and Problem~\ref{prob:SER} are equivalent.}.

\begin{Prob}[SER Minimization] \label{prob:SER}
\begin{eqnarray}
&\underset{\hat{\mathbf{x}}(t)}{\textrm{minimize}} & \lim_{N \to \infty }\sum\limits_{n=1}^N \sum\limits_{t=(n-1)T+1}^{nT}\frac{\mathcal{S}\left(\mathbf{x}(t) - \hat{\mathbf{x}}(t)\right)}{N\times T\times N_t} \\
& \textrm{subject to} &\mathbf{y}(t) = \mathbf{H}_n \mathbf{x}(t) + \mathbf{n}(t),\nonumber \\
&& \forall t\in \left[(n-1)T+1,nT\right], \\
&& \widehat{\mathbf{H}}_n = \mathbf{H}_n + \Delta \mathbf{H}, \quad \widehat{\mathbf{H}}= \left\{\widehat{\mathbf{H}}_n\right\}, \\
&& \hat{\mathbf{x}}(t) = \arg \min_{\hat{\mathbf{x}}} g\left(\mathbf{y}(t), \widehat{\mathbf{H}}\right).
\end{eqnarray}
where $\mathcal{S} (\cdot)$ calculates the number of non-zero elements in the inner vector.
\end{Prob}

By configuring different objective functions of $g(\cdot)$, traditional detection algorithms minimize the SER results through maximizing the likelihood function \cite{yecsilyurt2017hybrid}, minimizing the MSE \cite{MMSE} or zero-forcing the noise values \cite{li2018accurate}, where the mathematical expressions are\footnote{$(\cdot)^{H}$ denotes the matrix transpose and conjugation operation.},
\begin{eqnarray}
\hat{\mathbf{x}}(t)_{ML} & = & \arg\min\limits_{\mathbf{x}(t)}\parallel \mathbf{y}(t)-\widehat{\mathbf{H}}\mathbf{x}(t) \parallel^{2}, \\
\hat{\mathbf{x}}(t)_{ZF} & = & (\widehat{\mathbf{H}}^{H}\widehat{\mathbf{H}})^{-1}\widehat{\mathbf{H}}^{H}\mathbf{y}(t), \\
\hat{\mathbf{x}}(t)_{MMSE} & = & (\widehat{\mathbf{H}}^{H}\widehat{\mathbf{H}}+\sigma_{n}^{2}\mathbf{I}_{N_t})^{-1}\widehat{\mathbf{H}}^{H}\mathbf{y}(t).
\end{eqnarray}

With the imperfect CSI (ICSI) model, the received signal vector $\mathbf{y}(t)$ can be written as,
\begin{equation}
\mathbf{y}(t) = (\widehat{\mathbf{H}} - \Delta \mathbf{H}) \mathbf{x}(t) + \mathbf{n}(t)=\hat{\mathbf{H}}\mathbf{x}(t) + \Bar{\mathbf{n}}(t).
\end{equation}
Define $\Bar{\mathbf{n}}(t) \triangleq \mathbf{n}(t)-\Delta \mathbf{H} \mathbf{x}(t)$ to be the equivalent noise with the channel estimation errors, and the equivalent noise variance is given by,
\begin{equation}
\sigma_{\Bar{\mathbf{n}}(t)}^{2}=\frac{1}{N_r}\textrm{Tr}\left\{\mathbb{E}\left[\Bar{\mathbf{n}}(t)\Bar{\mathbf{n}}(t)^H\right]\right\}=N_t\sigma_{e}^2+\sigma_{n}^2,
\end{equation}
where $\textrm{Tr}(\cdot)$ denotes the matrix trace operation. Since the equivalent noise variance scales with the estimation error $\sigma_{e}^2$, the post-processing Signal-to-Noise Ratio (SNR) as well as the BER scales with the order $\mathcal{O}\left((\sigma_{e}^2)^{-1}\right)$.

\section{Neural Network Frame} \label{sect:frame}
Due to the non-convex approximation capability provided by neural networks, we can directly solve problem~\ref{prob:BER} with sufficient training data. In addition, since the neural networks provide outstanding generalization capability, the BER performance under imperfect channel conditions can be improved with the deep learning framework. As problem~\ref{prob:BER} defines an end-to-end BER evaluation framework under imperfect channel knowledge with binary decision of $\hat{\mathbf{b}}(t)$, it belongs to the classification problem in the machine learning area \cite{yan2017signal} and the loss function that commonly adopted is the cross-entropy between the estimation $\hat{\mathbf{b}}(t)$ and the original bits $\mathbf{b}(t)$, i.e.,
\begin{eqnarray} \label{eqn:loss}
\mathcal{L} = -\frac{\sum\limits_{n=1}^N \sum\limits_{t=(n-1)T+1}^{nT}\left[\mathbb{CE}\left(\hat{\mathbf{b}}(t), \mathbf{b}(t)\right)\right]}{M \times N_t \times NT},
\end{eqnarray}
where $\mathbb{CE}(a, b)$ is defined to be $ a \ln b + (1 - a) \ln (1 - b)$. 

\subsection{Data Preprocessing}
We use the MIMO channel model we have mentioned in Section~\ref{sect:system} and get the complex time-domain vectors of both $\mathbf{y}(t)$ and $\hat{\mathbf{H}}$, then convert them into real domain vectors. The details of the transition can be found in the following equation:
\begin{eqnarray}
\nonumber\dot{\mathbf{y}}(t) &=&{\left[ \begin{array}{cc}
\Re(\mathbf{y}(t))\\
\Im(\mathbf{y}(t))
\end{array}
\right]},
\dot{\mathbf{x}}(t) ={\left[ \begin{array}{cc}
\Re(\mathbf{x}(t))\\
\Im(\mathbf{x}(t))
\end{array}
\right]},\\
\dot{\mathbf{H}} &=&{\left[ \begin{array}{cc}
\Re(\hat{\mathbf{H}})&-\Im(\hat{\mathbf{H}})\\
\Im(\hat{\mathbf{H}})&\Re(\hat{\mathbf{H}})
\end{array}
\right]}.
\end{eqnarray}
where $\Re(\cdot)$ and $\Im(\cdot)$ are the real and imaginary part of complex vector respectively. 

\subsection{Network Architecture}
To find a network structure suitable for MIMO detection, we have tested two neural networks, CNN and DNN. DNN is often used for joint channel equalization and decoding, while CNN is widely applied to do feature extraction and process correlated noise \cite{unified18}. The layout of these two networks we used are given in the following Table~\ref{net_par}. For DNN, there are four hidden layers, besides the output layer, input layer and a Batch Normalization layer used to accelerate convergence and prevent overfitting. And for CNN, two convolutional layers are followed by two dense layers. 
 
\begin{figure}[htbp]
\centering
        \includegraphics[width=3in]{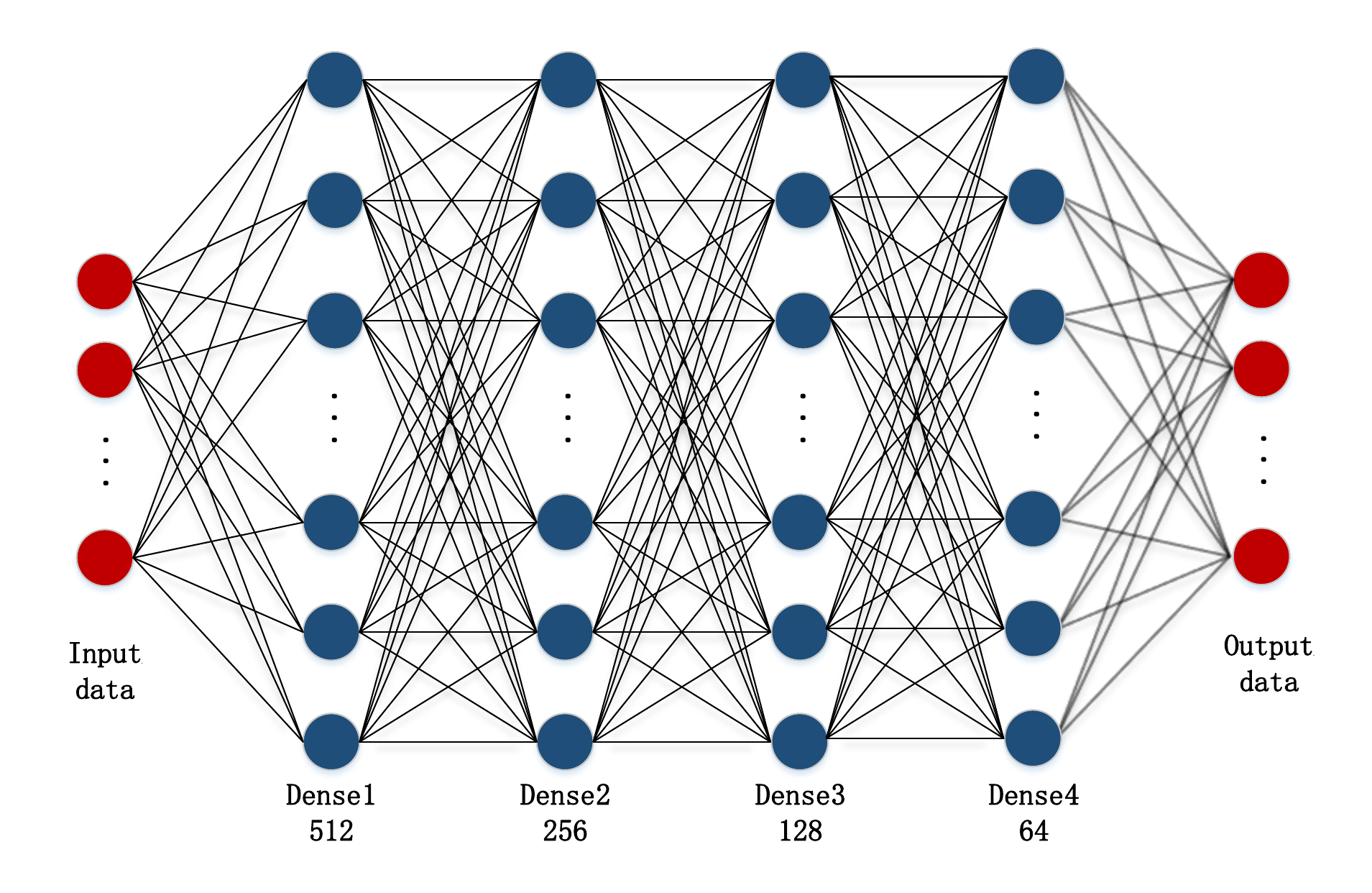}
            \caption{Architecture of the DNN model used in MIMO detection.}
        \label{fig:DNN}
\end{figure}

\begin{figure}[htbp]
\centering
        \includegraphics[width=3.3in]{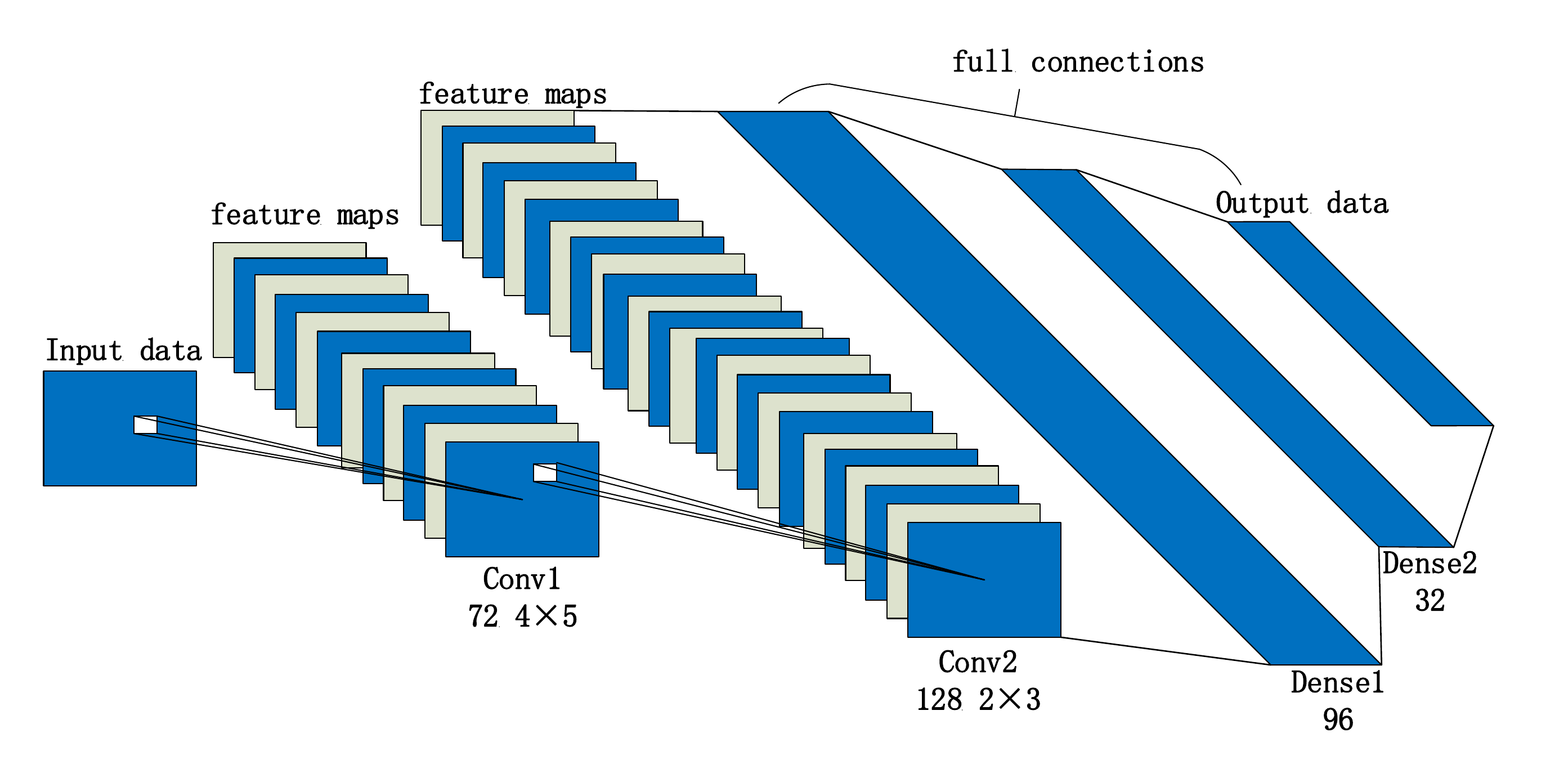}
            \caption{Architecture of the CNN model used in MIMO detection.}
        \label{fig:CNN}
\end{figure}

 The architectures of DNN and CNN model are shown in Fig.~\ref{fig:DNN} and Fig.~\ref{fig:CNN} respectively. The first parameter below each convolutional layer represents the number of filters in that layer, while the second and third numbers show the size of each filter. For the two dense layers, there are 128, 64 neurons.
\begin{table}
\caption{An Overview of Network Configurations and Parameters. \label{net_par}}
\centering
\footnotesize
\setlength{\tabcolsep}{1.0mm}{
\begin{tabular}{c c c}
\toprule
           &DNN  &CNN\\
\midrule
\midrule
Input Layer   & 4*5/8*9         & 8*9*1\\
\midrule
Layer1     &Dense 512-ReLU       & 72*4*5-ReLU\\
\midrule
Layer2     &BatchNormalization    &BatchNormalization\\
\midrule
Layer3     &Dense 256-ReLU        &128*2*3-ReLU\\
\midrule
Layer4     &Dense 128-ReLU        &Dense 96-ReLU\\  
\midrule
Layer5     &Dense 64-ReLU        &Dense 32-ReLU\\
\midrule
Output Layer & 4-way sigmoid   & 4-way sigmoid\\
\midrule
Total Parameters & 215372  &208012\\
\bottomrule
\end{tabular}}
\end{table}

\begin{itemize}
\item{\em Input Layer:} Joint matrix of $\dot{\mathbf{y}}(t)$ and $\dot{\mathbf{H}}$ which have been converted to real domain from complex domain. 
\item{\em Hidden Layers:} Deploy neural network to estimate the mapping function: $h\left(\mathbf{y}(t), \widehat{\mathbf{H}}\right)$ and the activation function is {\em rectified linear unit} ({\em ReLU}).
\item{\em Output Layer:} $\hat{\mathbf{b}}(t)$, and the activation function is {\em sigmoid}, because we can take this problem as a classification problem.
\item{\em Optimizer:} We choose to use {\em adam} as the optimizer which is widely used in deep learning.
\item{\em Loss Function:} Our problem is a discrete Ont-Hot vector of a classification problem. So we used cross entropy as our loss function.
\end{itemize}

\subsection{Detection Performance of DNN versus CNN}
\begin{figure}[htbp]
\centering
        \includegraphics[width=2.5in]{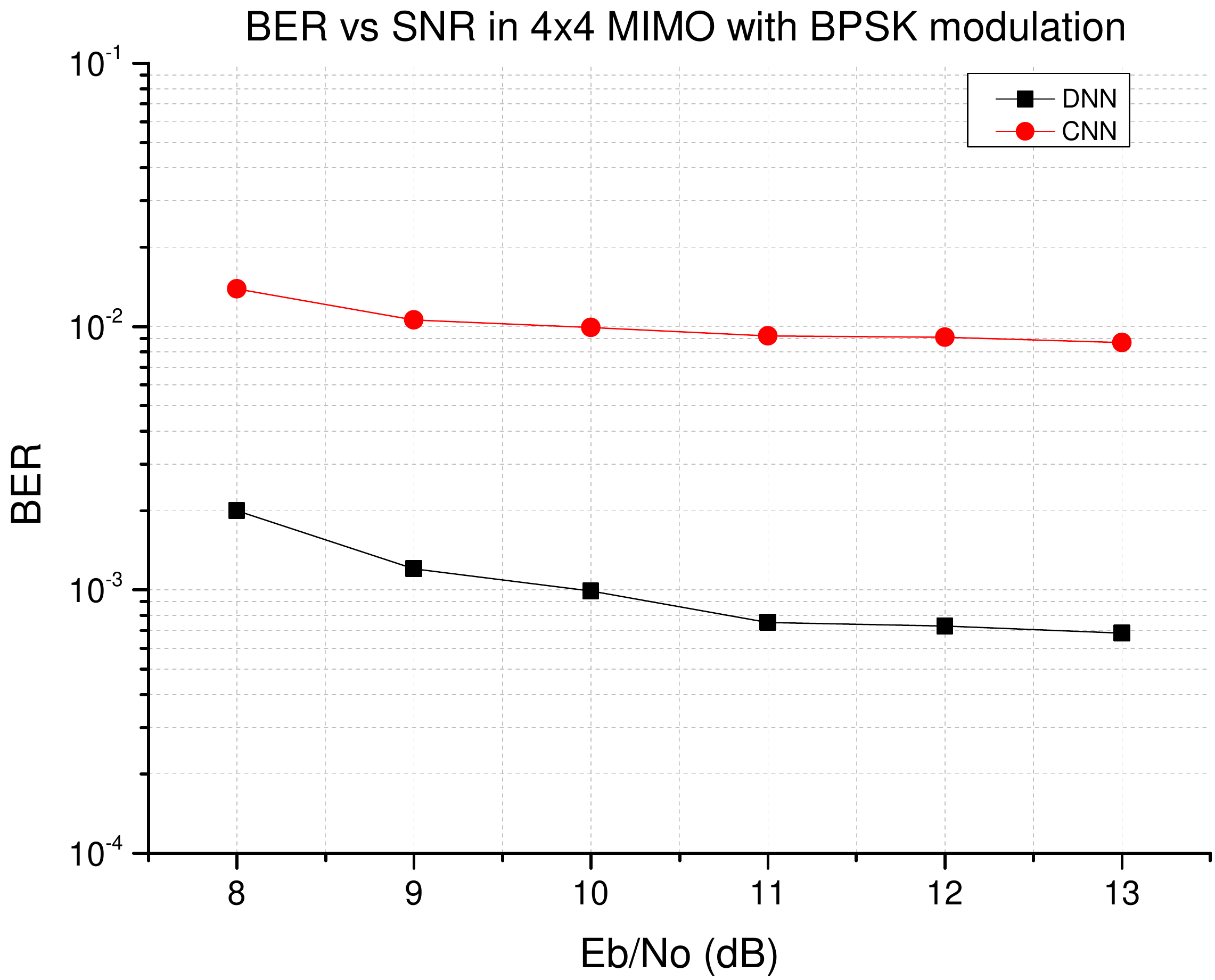}
            \caption{The BER performance of CNN versus DNN with perfect CSI. In this experiment, DNN-based detection method performs much better CNN-based method.}
        \label{fig:CNNDNN}
\end{figure}

In order to test which network is more suitable for MIMO detection, we apply these two networks with a close number of total parameters to fairly perform $4\times4$ MIMO detection with BPSK modulation. The details of the experiment are given in Section~\ref{sect:experiment}. The result is shown in Fig.~\ref{fig:CNNDNN}. It proves that DNN has better BER performance than CNN. We believe this is on account of the strong ability of DNN to process one-dimensional data. Besides, we have measured the run time to detect $7.2\times 10^5$ symbols needed by these two neural network. On the one hand, it took DNN 15 seconds to finish the detection. On the other hand, CNN took 58 seconds, which is nearly 4 times that of DNN. In summary, DNN is more suitable for MIMO detection in term of BER performance and decoding rate.

\section{Simulation Results} \label{sect:experiment}
We have compared the DNN-based method with ZF (Baseline 1), MMSE (Baseline 2), DetNet \cite{DeepMIMOdetetion} (Baseline 3) and ML-based detection algorithm (Baseline 4) in term of BER performance, robustness and decoding rate. The parameters of experiments are shown in Table~\ref{sim_par}. Besides, we consider the spatial correlation mentioned in document \cite{saleh2009performance}. We assume that the fading of the receiver is spatially uncorrelated but transmit-correlated, which is typical in the downlink channel of a mobile communication system.
\begin{table}
\caption{Simulation Parameters. \label{sim_par}}
\centering
\footnotesize
\setlength{\tabcolsep}{1.0mm}{
\begin{tabular}{c c c}
\toprule
           &parameters\\
\midrule
\midrule
computer setup     &Intel core i5-6500 CPU @3.20GHz with a 12GB memory\\
\midrule
training set     &$5.4 \times 10^5$ symbols\\
\midrule
test set     &$7.2 \times 10^5$ symbols\\
\midrule
validation set   &$1.8 \times 10^5$ symbols\\
\bottomrule
\end{tabular}}
\end{table}

\subsection{BER Performance with Perfect CSI}\label{sect:perfect}
\begin{figure}[htbp]
\centering
        \includegraphics[width=3in]{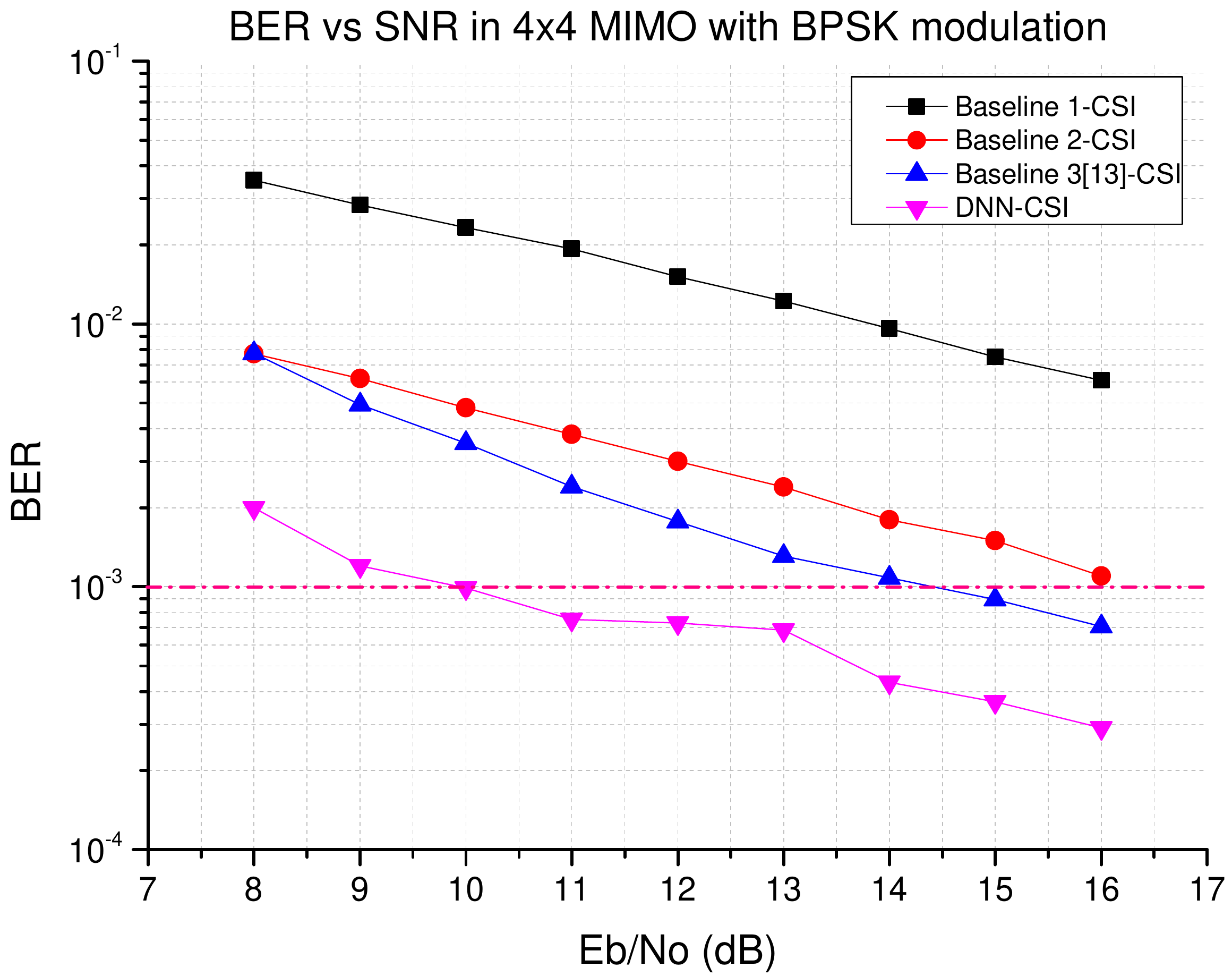}
            \caption{BER versus SNR with perfect CSI. In this experiment, DNN-based detection method outperforms ZF, MMSE and DetNet.}
        \label{fig:p44}
\end{figure}

In this section, DNN-based approach is compared with DetNet and traditional methods, such as ZF and MMSE under different SNRs with full knowledge of the CSI (${N_p \cdot E_p}\to +\infty$, i.e. $\hat{\mathbf{H}}=\mathbf{H}$). In all the simulations, our networks are trained on data sets generated at 8dB SNR. The performance for $4\times4$ MIMO detection with BPSK modulation is evaluated and the result is given in Fig.~\ref{fig:p44}. As we can see from the figure, the BERs of DNN-based method are lower than MMSE and ZF at various SNRs, and even has an advantage of more than 5dB. So our network has a strong capability of generalization. Besides, for BER of $10^{-3}$, the DNN-based method outperforms DetNet in our channel model for 4.5 dB.

\subsection{Robustness with Imperfect CSI}\label{sect:imperfect}
\begin{figure}
\centering
\subfigure[BER versus SNR in $2\time2$ MIMO with QPSK modulation.]{
\begin{minipage}[b]{0.4\textwidth}
\includegraphics[width=1\textwidth]{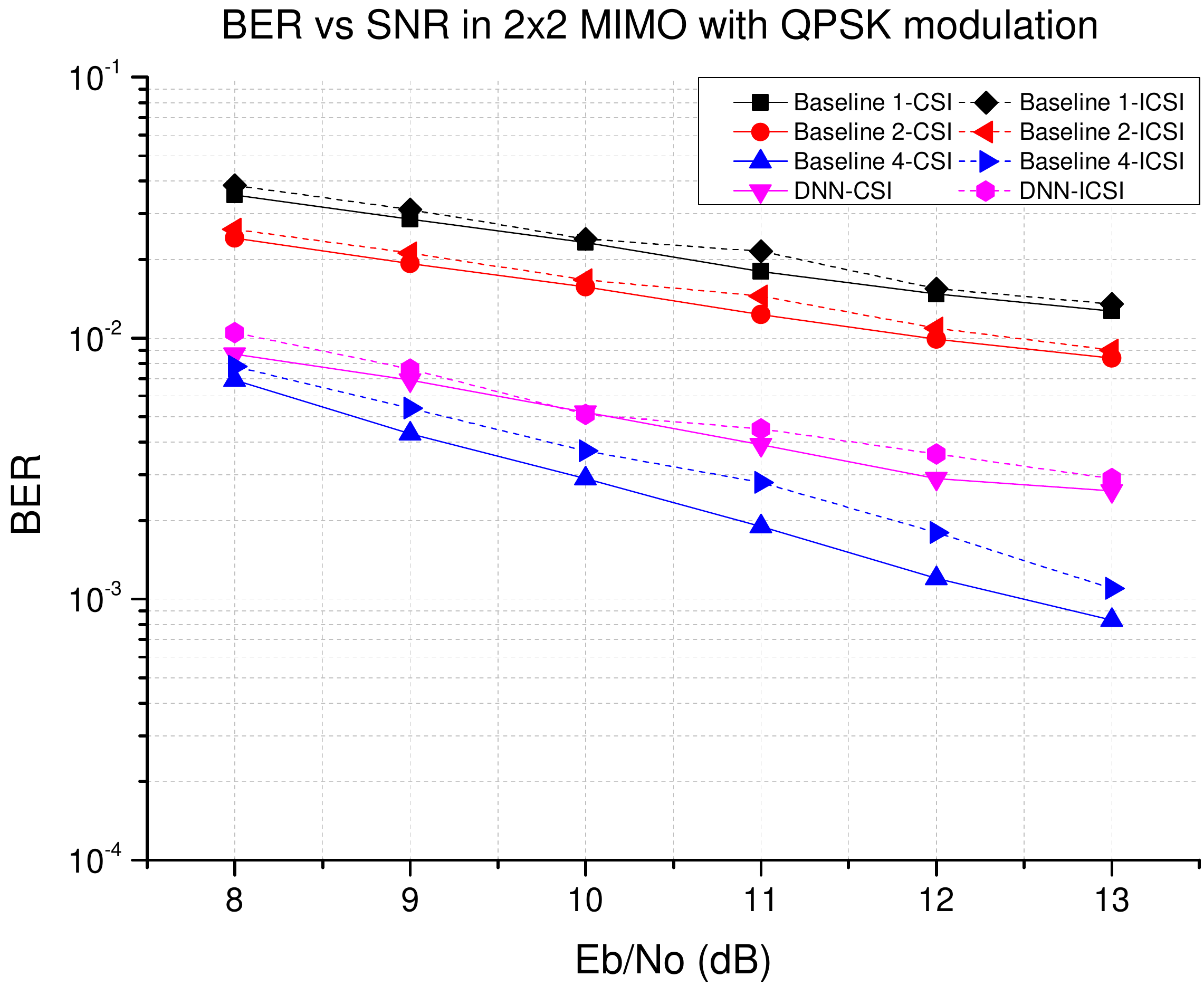}
\end{minipage}
}
\subfigure[BER versus SNR in $4\time4$ MIMO with BPSK modulation.]{
\begin{minipage}[b]{0.4\textwidth}
\includegraphics[width=1\textwidth]{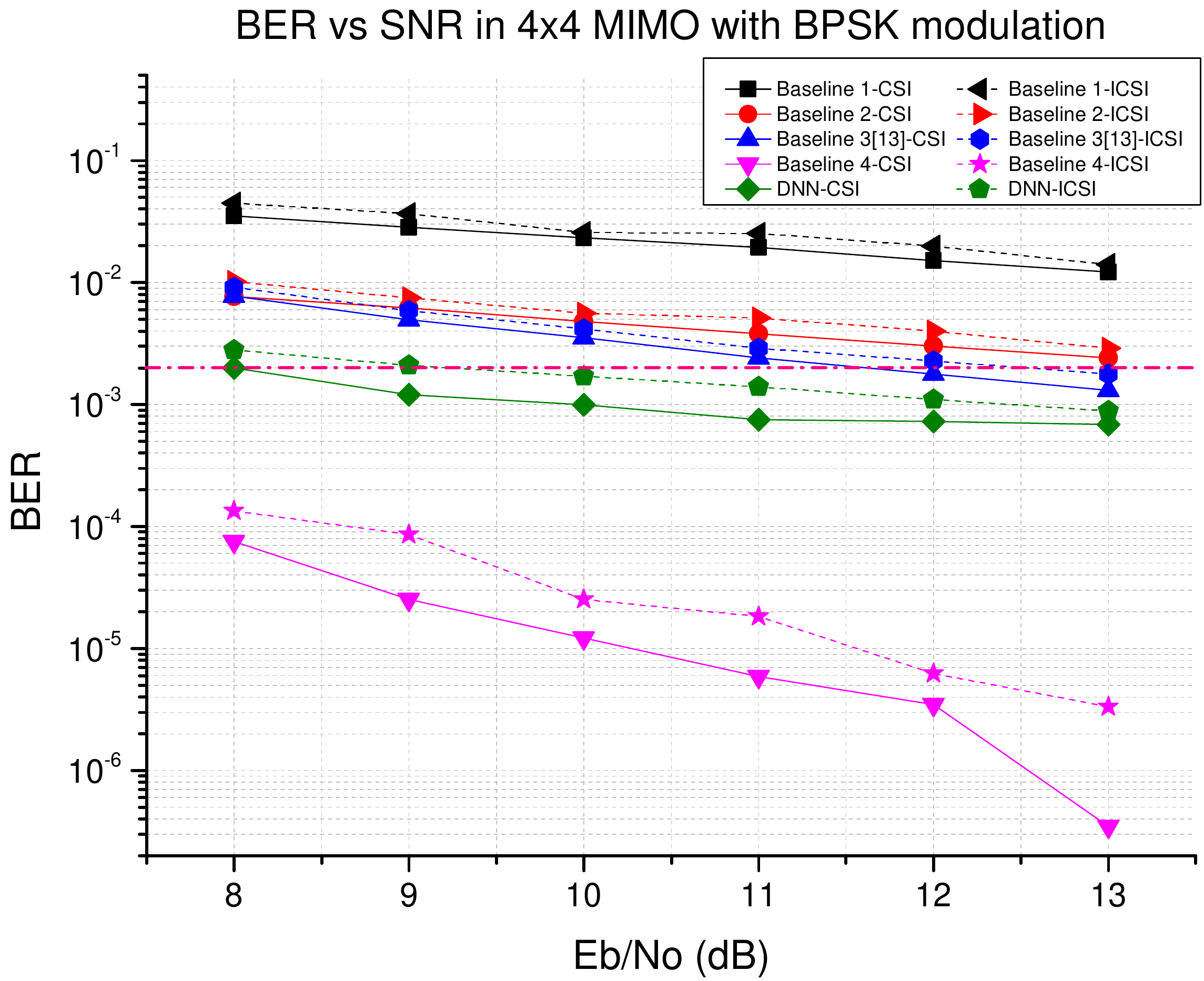}
\end{minipage}
}
 \caption{BER versus SNR with perfect and imperfect channel information. In this experiment, DNN-based detection method still outperforms MMSE, ZF and DetNet.} \label{fig:imperfect}
\end{figure}

The proposed method is compared with traditional methods, including ML with imperfect channel information in this section. We set $N_p \cdot E_p=400$, i.e., $\Delta \mathbf{H} \sim \mathbb{CN}(\mathbf{0},\frac{N_t}{400}\mathbf{I}_{N_r})$. The result for $2\times2$ MIMO detection with QPSK modulation and $4\times4$ MIMO detection with BPSK modulation are shown in Fig.~\ref{fig:imperfect}. We noticed that with imperfect channel knowledge, for BER of $10^{-2}$, the proposed DNN-based method still outperforms the MMSE and ZF methods for about 4.5 dB and more than 5dB when performing $2 \times 2$ MIMO detection with QPSK. Particularly, for $4 \times 4$ MIMO detection with BPSK, DNN-based approach outperforms the DetNet for about 3.5 dB for BER of $2\times 10^{-3}$, ZF and MMSE for more than 4 dB respectively. This indicates that the proposed method still has a good robustness with imperfect CSI. Moreover, we can notice that the BER of the ML method in the case of imperfect CSI are significantly higher compared with that of perfect CSI. But the fluctuation of DNN caused by the perfection of CSI is similar to other methods. It is worth mentioning that the detecting result of DNN-based method using imperfect CSI is even better than that of DetNet MIMO detection method with perfect channel information. Although the BER of the proposed method is higher than that of the ML method, using the  model for MIMO detection has much lower complexity than ML algorithm, which is enough to make up for the deficiency in BER. 

\subsection{Throughput Comparison among Different Method }
The higher time complexity the algorithm has, the slower the decoding rate is. To compare the time complexity of different detection algorithms, we calculate the average throughput of them, which equals to the number of detected bits divided by the time consumption. We program and run all MIMO detection algorithms in Python 3.5.2 using an Intel core i5-6500 CPU @3.20GHz with a 12GB memory. We run the test to detect $7.2 \times 10^5$ symbols and record the time consumption for 3 times, then calculate the average throughput. In Table \ref{com_with_ondeep}, we compare the detection efficiency of different schemes, including ZF, MMSE, ML and the proposed DNN-based detection method for $4\times4$ MIMO detection with BPSK modulation. From Table \ref{com_with_ondeep}, we can see the throughput of ZF is highest, and the proposed DNN-based method has very similar performance with ZF algorithm. MMSE followed and ML detection method is lowest. We can conclude that DNN-based method has a near-ZF throughput performance and achieve more accurate decoding performance at the same time, which means that the time complexity of DNN-based method is also low. 

\begin{table*}[htbp]
\caption{Throughput Comparison among Different Schemes}
\centering
\begin{tabular}{c c c c c c}
\toprule
                     &ZF          &MMSE            & ML             & DNN-based        \\
\midrule
Throughput (Kbps)    & $4.8196 \times 10^{4}$     & $4.6933 \times 10^{4}$      & $8.8377 \times 10^{3}$      & $4.8 \times 10^{4}$\\
\bottomrule
\end{tabular}
\label{com_with_ondeep}
\end{table*}

\section{Conclusion} \label{sect:conc}
In this paper, we compared two MIMO detection methods based on deep learning. The result shows that DNN has better BER performance and higher decoding rate than CNN. We performed simulation experiments in the case of perfect and imperfect channel information respectively, and compared the performance of the proposed method with the traditional method and DetNet under different SNR conditions. In the case of all SNRs, DNN-based method has a certain distance from the ML method, but better than the other detection schemes. In addition, it has similar decoding rate with ZF method, which far exceeds ML method. Besides, better BER performance and robustness can be achieved in the case of imperfect channel information, compared with the ZF, MMSE methods and DetNet. Furthermore, the proposed network is easy to be implemented on hardware owing to its simpler network structure. In general, DNN-based method is a proper approach for MIMO detection.

\section*{Acknowledgement}
This work was supported by the National Natural Science Foundation of China (NSFC) Grants under No. 61701293, the National Science and Technology Major Project Grants under No. 2018ZX03001009, the Huawei Innovation Research Program (HIRP), and research funds from Shanghai Institute for Advanced Communication and Data Science (SICS).

\bibliographystyle{IEEEtran}
\bibliography{IEEEfull,MIMOdetection}

\end{document}